\documentclass[prd,preprint,showpacs,letterpaper]{revtex4}
\usepackage{epsfig}
\usepackage{changebar}
\usepackage{dcolumn}
\usepackage{amsmath}
\begin{document}
\preprint{}
\title{Sensitivity to the KARMEN Timing Anomaly at MiniBooNE}
\author{S. Case}
  \affiliation{Columbia University, New York, NY 10027}
\author{S. Koutsoliotas} 
  \affiliation{Bucknell University, Lewisburg, PA 17837}
\author{M. L. Novak}
  \affiliation{Bucknell University, Lewisburg, PA 17837}
\date{\today}

\begin{abstract}
We present sensitivities for the MiniBooNE experiment to
a rare exotic pion decay producing a massive particle, $Q^0$.  This 
type of decay represents one possible 
explanation for the timing anomaly reported by the KARMEN collaboration.
MiniBooNE will be able to explore an area of the KARMEN
signal that has not yet been investigated. 
\end{abstract}

\maketitle
\section*{The KARMEN Timing Anomaly}

The KARMEN collaboration at the ISIS spallation neutron 
facility at the Rutherford Appleton Laboratory uses a pulsed 
neutrino beam from stopped pion and muon decays to study 
neutrino-nucleon interactions.  They have reported an anomaly 
in the timing distribution of neutrino interactions from 
stopped muon decays \cite{Katime}.  One possible 
explanation for the anomaly 
is an exotic pion decay where a neutral and weakly interacting 
or sterile particle ($Q^0$) with a velocity of 4.9 $m/ \mu s$ is produced.  
This particle, with a mass of 33.9 $MeV / c^2$ measured from the time 
of flight,  then decays in the KARMEN detector to $e^+ e^- \nu$ 
or $\gamma \nu$.  The decay to $e^+ e^- \nu$ is strongly favored 
by recent data \cite{Kaeenu}.

The KARMEN experiment reports a signal curve for pion branching 
ratio B($\pi \rightarrow \mu Q^0$) $\times$ B($Q^0 \rightarrow$ visible) 
versus lifetime.  A minimum branching ratio of $10^{-16}$ exists
for a lifetime of \hbox{3.6 $\mu s$}.  Two solutions exist above 
this minimum, one at large and one at small lifetimes.  
Experiments at the Paul Scherrer Institute (PSI) have excluded 
any exotic pion decays to muons with branching ratios above 
$6.0 \times 10^{-10}$ at 90\% CL, and to electrons with branching ratios above 
$9.0 \times 10^{-7}$ at 90\% CL \cite{PSIlett,PSIrev,PSInew}.
Also, the NuTeV experiment (E815) at Fermi 
National Accelerator Laboratory (Fermilab) has excluded the short 
lifetime solution with branching ratios above $5 \times 10^{-12}$ 
at 90\% CL \cite{NuTeV}.
In addition, astrophysical constraints on certain decay modes 
of the $Q^0$ particle exclude lifetimes greater than $10^3$ 
seconds.  Still, segments of the KARMEN signal region remain 
to be addressed.

Although recent data from KARMEN2 does not show a statistically significant 
signal for the $Q^0$ particle, a substantial effect persists in the 
combined data sets from KARMEN1 and KARMEN2 \cite{Karmen2}.

\section*{The MiniBooNE experiment}

The MiniBooNE experiment (E898), currently under construction at Fermilab, 
will begin collecting data in 2002. The Fermilab Booster ring will 
provide 8 $GeV$ protons, which will strike a beryllium target.  
This interaction will produce primarily 
pions, which will decay in flight in a 50 $m$ decay pipe.  
Muons from the pion decays will be stopped in a steel beam dump
and about 489 $m$ of earth 
between the decay pipe and the center of the MiniBooNE detector.  
The MiniBooNE detector 
is a spherical tank, 6.1 $m$ in radius, and  filled with 807 $tons$ of 
mineral oil.  An inner tank structure at 5.75 $m$ radius supports 
1280 8-inch photomultiplier tubes facing into the tank.  An additional 
330 8-inch photomultiplier tubes in the optically isolated outer region 
of the tank will serve as a veto.  The MiniBooNE 
experiment will receive approximately $5 \times 10^{20}$ 
protons on target each year.

\begin{figure}
\centering
\epsfig{file=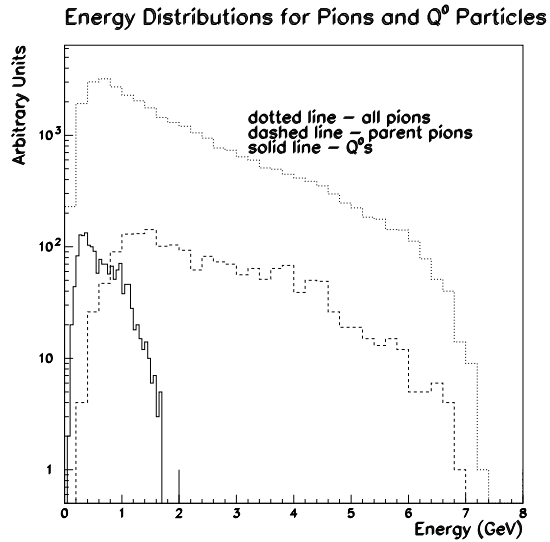,height=3in}
\caption{Energy distributions for particles in the Monte Carlo model.  
The distribution for all pions produced, the pions which decay to 
$Q^0 s$ within the detector acceptance, and the $Q^0$ energy spectrum are shown.}
\label{energy}
\end{figure}

A GEANT-based \cite{Monte} Monte Carlo was used to simulate the 
production and decay of the $Q^0$ particles from pion decay using the decay mode 
\hbox{$Q^0 \rightarrow e^+ e^- \nu$}. Figure \ref{energy} shows the energy 
spectrum of all pions produced in our model. ``Parent pions'' which 
produce $Q^0 s$ within the detector acceptance, and the $Q^0$ particles 
that reach our detector are also shown.  The number of $Q^0 s$ decaying in the 
MiniBooNE detector in a year of running was then calculated for a variety of 
lifetimes and branching ratios within the KARMEN signal band.
The electron and the positron will represent the signal signature and will 
be reconstructed from their \v{C}erenkov
rings in the detector.  In most cases, the opening angle between the 
electron and the positron appears to be too small to resolve into two separate 
\v{C}erenkov rings.  Consequently, the major background to the 
$Q^0$ signal will be $\nu_e$ quasielastic scattering. In the cases where the
two rings are separately resolvable, neutral current $\pi^0$ production will be the 
dominant background contribution.

\section*{The Booster Timing Structure}

\begin{figure}
\centering
\epsfig{file=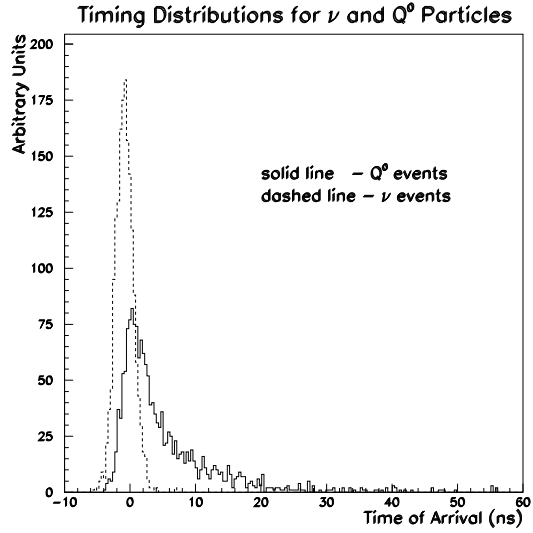,height=3in}
\caption{Timing distribution for events at the MiniBooNE detector.  
The solid line represents the $Q^0$ particles, and the dashed line shows the 
distribution of neutrinos from the beam.}
\label{1ns}
\end{figure}

\begin{figure}
\centering
\epsfig{file=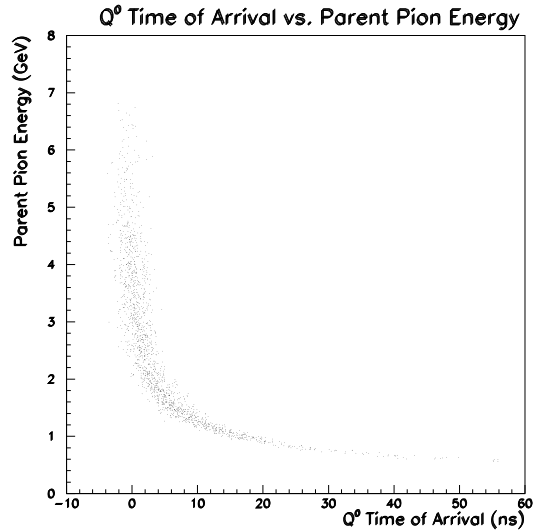,height=3in}
\caption{The $Q^0$ time of arrival at the MiniBooNE detector 
as a function of the energy of the parent pion.}
\label{scatter}
\end{figure}

\begin{figure}
\centering
\epsfig{file=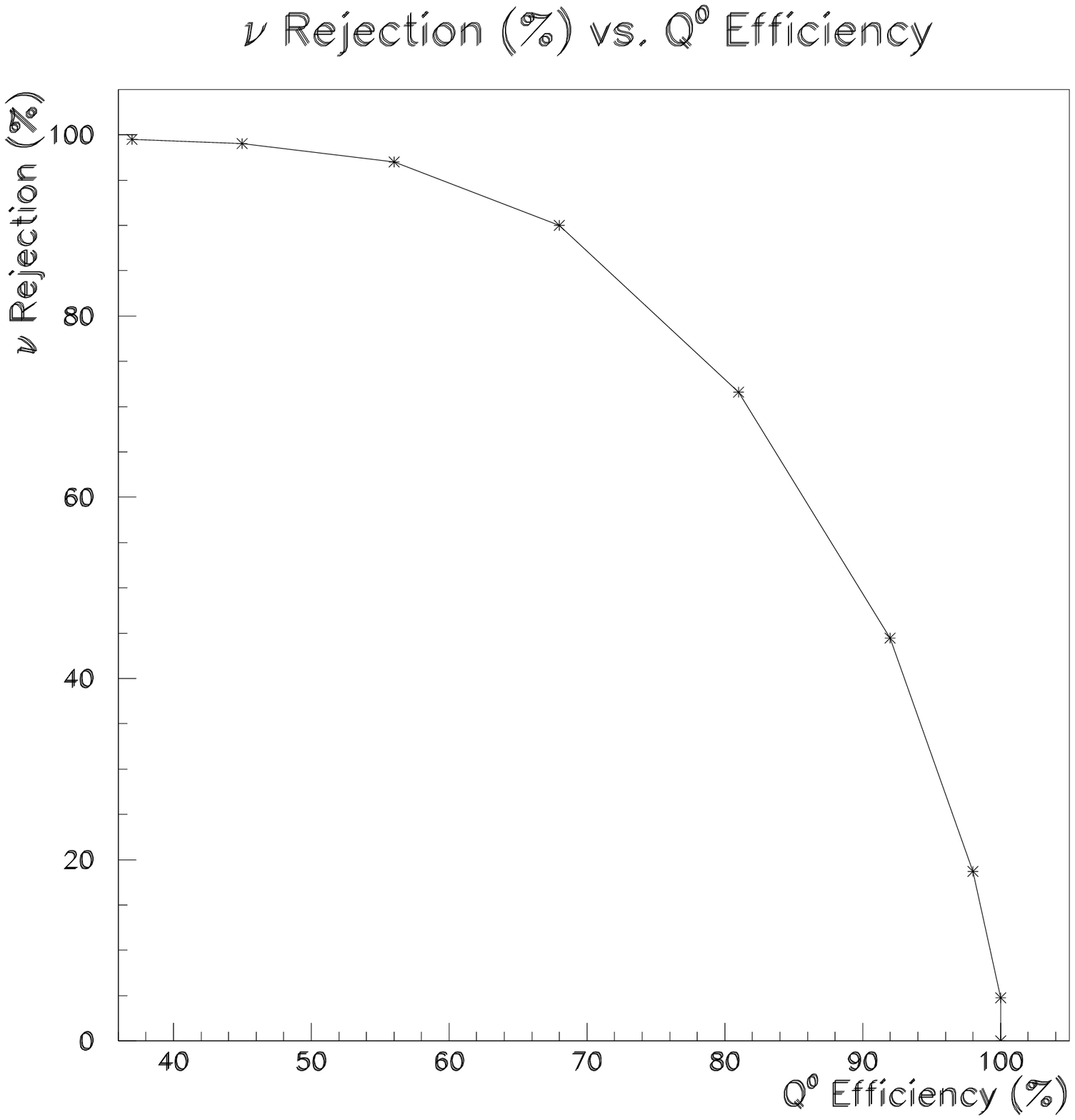,height=3in}
\caption{The $\nu$ rejection (\%) vs. $Q^0$ efficiency for 
different timing cut values.}
\label{rvse}
\end{figure}

The timing structure of the proton pulses from the Booster 
allows us to discriminate between neutrino and $Q^0$ events.  
Micropulses within the Booster pulses can be 
approximated by gaussian-shaped pulses with widths of 1.5 $ns$, 
separated by 18.94 $ns$.  
Given that the $Q^0$ mass is very close to the kinematic limit for the 
decay $\pi \rightarrow \mu Q^0$, the $Q^0$ particles will be traveling 
at speeds similar to the parent pions, and considerably slower than the
neutrinos.  Most of the neutrinos will arrive at the detector within 
a tight timing bunch, while many of the $Q^0 s$ will reach the 
detector at times between the neutrino bunches. 
Figure \ref{1ns} shows the timing 
distributions for a neutrino bunch and for the corresponding $Q^0$ 
particles, and Figure \ref{scatter} shows the arrival time of the 
$Q^0$ particle as a function of the energy of the parent particle.  The timing 
structure makes it possible to impose a cut on the time of arrival which is
 very effective in isolating the $Q^0$ signal. By varying the value of 
the timing  cut, the $Q^0$ detection efficiency was examined. 
 Figure \ref{rvse} shows the efficiency of detecting a $Q^0$ particle 
as a function of neutrino rejection.  We have assumed MiniBooNE's timing resolution
for neutrino events to be 1.2 $ns$, and independent of neutrino energy.

This paper presents sensitivities of the MiniBooNE experiment to 
the KARMEN timing anomaly based on $Q^0$ efficiencies of 10\% and 
100\%.  It is anticipated that the final MiniBooNE sensitivity will 
lie between these limits.  For this analysis we also assume no 
backgrounds to the $Q^0$ signal.  

\section*{MiniBooNE Sensitivity to the KARMEN Timing Anomaly}

\begin{figure}
\centering
\epsfig{file=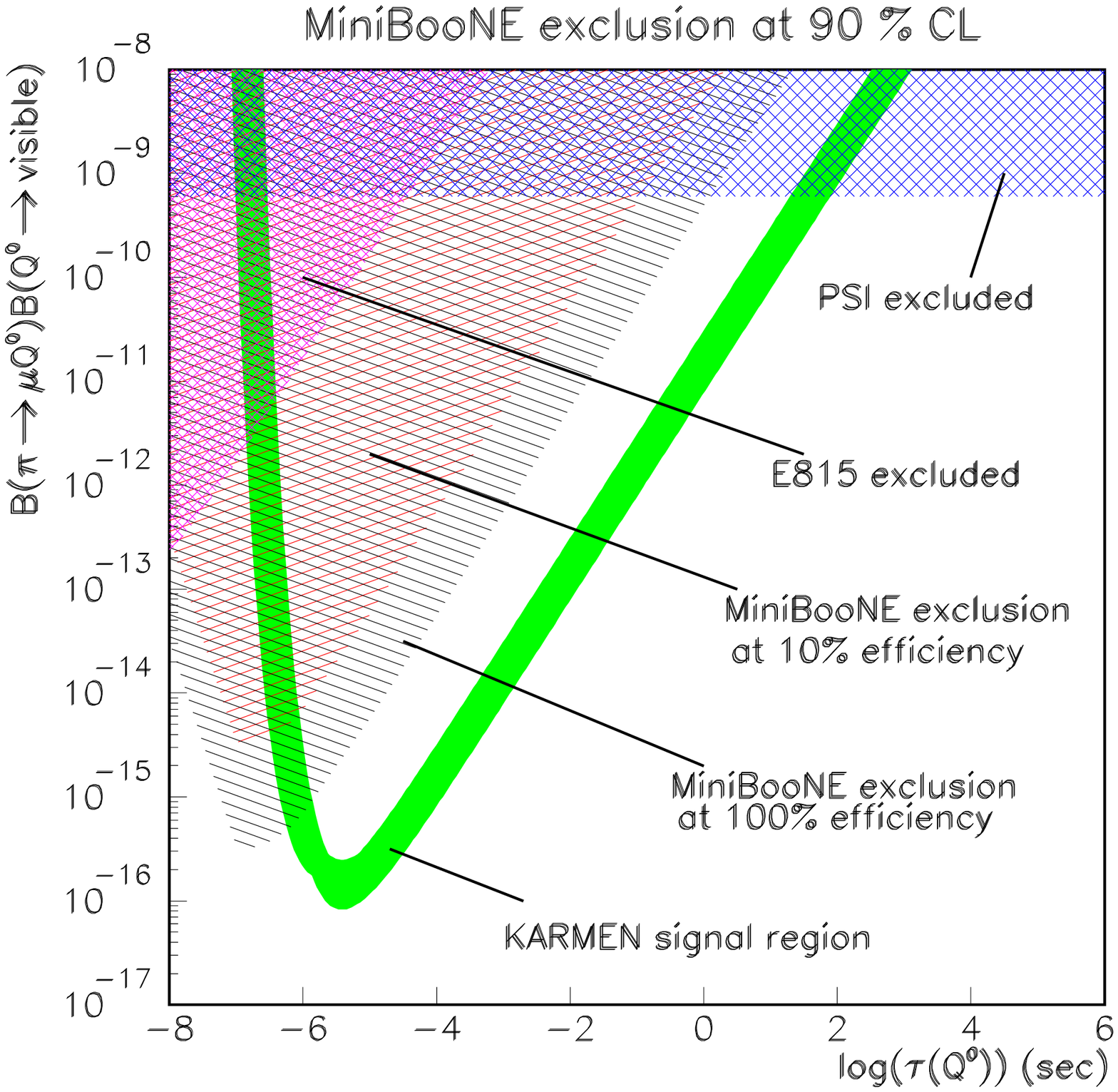,width=3.1in}
\caption{MiniBooNE's exclusion reach at 90\% CL based on one year of 
running and assuming no background.}
\label{exclude}
\end{figure}

MiniBooNE will be able to explore the short-lifetime region of the 
KARMEN signal at 90\% CL down to branching ratios of approximately 
$5 \times 10^{-15}$ at 10\% detection efficiency. Figure \ref{exclude}
shows MiniBooNE's exclusion region for one year of running and assuming
no background.  MiniBooNE's maximum reach in the short-lifetime 
solution may be extended to a branching ratio of $5 \times 10^{-16}$ if
 100\% detection efficiency is assumed. 
These limits are based on a preliminary version of the reconstruction 
algorithm.  Improvements to the sensitivity are anticipated with further 
development of the reconstruction algorithm.

\begin{figure}
\centering
\epsfig{file=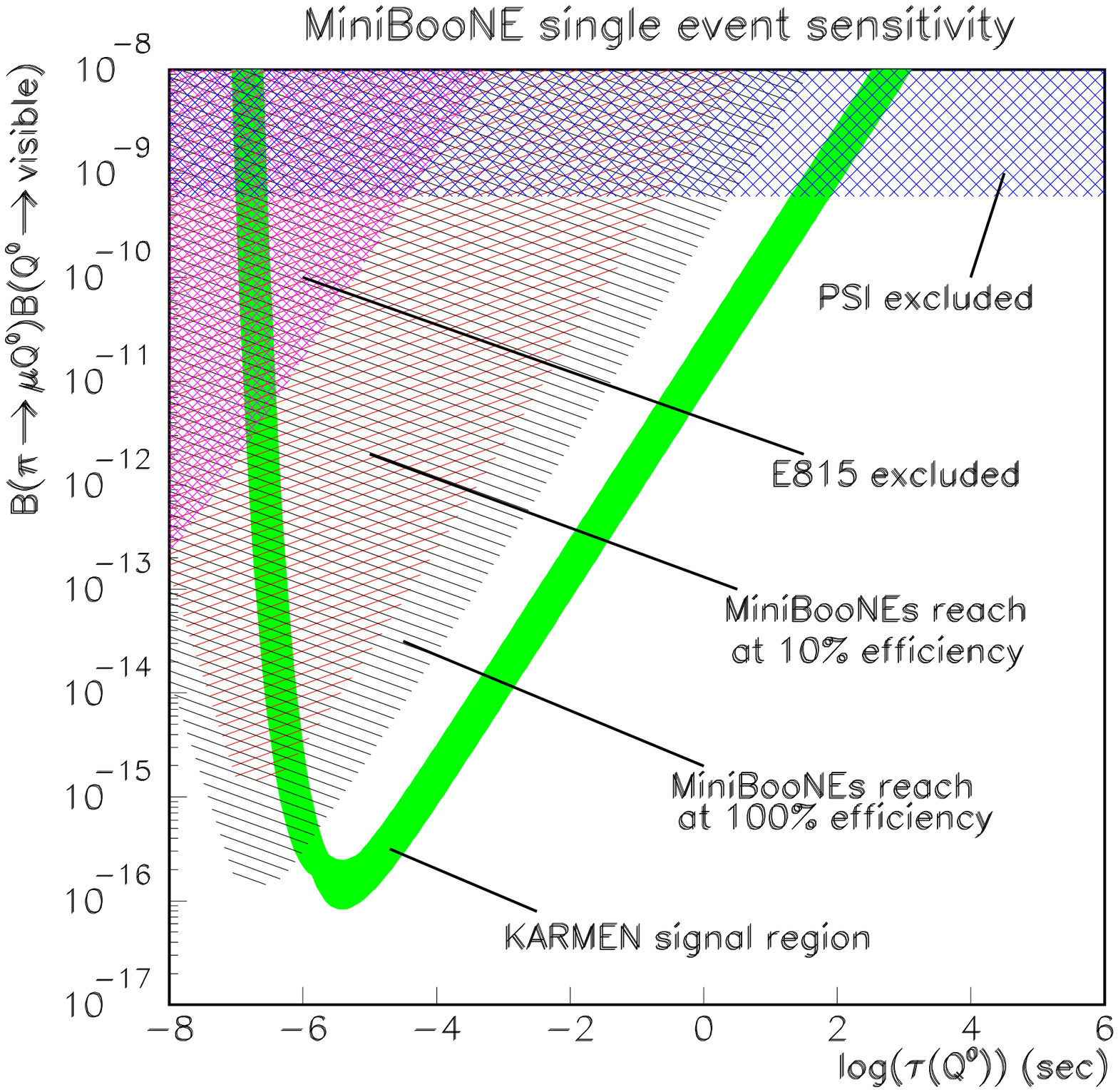,width=3.1in}
\caption{MiniBooNE's single-event sensitivity in one year of running, 
assuming no background.}
\label{event}
\end{figure}

Our single event sensitivity in the short-lifetime 
solution will reach branching ratios of approximately $2 \times 10^{-15}$ 
at 10\% detection efficiency  in one year of running, with no background.  
The maximum single-event sensitivity (based on 100\% detection efficiency)
 will reach branching ratios of approximately $2 \times 10^{-16}$ in one 
year of running with no background \hbox{(Figure \ref{event})}.

In conclusion, the timing anomaly observed by the KARMEN1 experiment has 
not been explained.
The MiniBooNE experiment will be capable of probing previously unexplored 
regions of the KARMEN signal.

\section*{Acknowledgments}
The authors thank the E898 (MiniBooNE) collaboration, with special 
thanks to J. Conrad, J. Formaggio, M. Shaevitz, and E. D. 
Zimmerman of Columbia University.  We also 
thank the National Science Foundation for their support.


\begin{thebibliography}{99}
\bibitem{Katime} B. Armbruster \textit{et al., Phys. Lett.} B 348, 19 (1995). 
\bibitem{Kaeenu} C. Oehler, \textit{Nucl. Phys.} B, Proc. Suppl. 85, 101 (2000). 
\bibitem{PSIlett} M. Daum \textit{et al., Phys. Lett.} B 361, 179 (1995). 
\bibitem{PSIrev} N. De Leener-Rosier \textit{et al., Phys. Rev.} D 43, 3611 (1991). 
\bibitem{PSInew} M. Daum \textit{et al., Phys. Rev. Lett.} 85, 1815 (2000). 
\bibitem{NuTeV} J. A. Formaggio \textit{et al., Phys. Rev. Lett.} 84, 4043 (2000).  
\bibitem{Karmen2} K. Eitel \textit{et al., Nucl. Phys.} B, Proc. Suppl. 91, 191 (2001); K. Eitel, private communication, July 2001; R. Schrock, private communication, July 2001.
\bibitem{Monte} Applications and Software Group, CERN, ``GEANT: Detector Description and Simulation Tool,'' CERN Program Library Report Q123.

\end{thebibliography}
\end{document}